\documentclass[aps, prd, preprint, superscriptaddress, tightenlines, nofootinbib]{revtex4}

\usepackage{amssymb, amsmath} 
\usepackage{stmaryrd}

\usepackage{bbm} 

\usepackage[english]{babel}
\usepackage{slashed}
\usepackage{cancel}

\usepackage{slashbox}

\usepackage{graphicx}
\usepackage{color}

\usepackage{datetime}
\settimeformat{ampmtime}

\usepackage{float}

\usepackage[bookmarks,breaklinks,plainpages=false,pdfpagelabels]{hyperref} 
	\hypersetup{linkcolor=red,citecolor=blue,filecolor=dullmagenta,urlcolor=darkblue} 

\definecolor{grey}{rgb}{0.4,0.4,0.4}
\definecolor{lightgrey}{rgb}{0.6,0.6,0.6}
\definecolor{dullmagenta}{rgb}{0.4,0,0.4}
\definecolor{darkblue}{rgb}{0,0,0.4}
\definecolor{orange}{rgb}{1,0.5,0}
\definecolor{lightbrown}{rgb}{0.75,0.5,0.25}
\definecolor{tan}{cmyk}{0.14,0.42,0.56,0}
\definecolor{djunglegreen}{cmyk}{0.99,0,0.52,0}
\definecolor{lightgreen}{rgb}{0,1,0}
\definecolor{olivegreen}{cmyk}{0.64,0,0.95,0.40}
\definecolor{midgreen}{rgb}{0.0,0.675,0.0}

\newcommand{\vs}{\vspace}

\renewcommand{\.}{\hspace{0.5mm}}

\newcommand{\ra}{\ensuremath{\rightarrow}}

\newcommand{\Ocal}{\ensuremath{\mathcal{O}}}

\renewcommand{\d}{\ensuremath{\mathrm{d}}}

\newcommand{\defas}{\mathrel{\mathop :}=} 

\newcommand{\eg}{e.g.}
\newcommand{\ie}{i.e.}

\newcommand{\cf}{cf.}

\makeatletter
 \def\ifundefined#1{\expandafter\ifx\csname#1\endcsname\relax}
 \ifundefined{default@color}
  \let\default@color=\current@color
 \fi
\makeatother

\newcommand{\beq}{\begin{equation}}
\newcommand{\eeq}{\end{equation}}
\newcommand{\bea}{\begin{eqnarray}}
\newcommand{\beas}{\begin{eqnarray*}}

\newcommand{\eea}{\end{eqnarray}}
\newcommand{\eeas}{\end{eqnarray*}}

\begin{document}

\title{Modified Bose-Einstein Condensate Black Holes\\ in $\boldsymbol{d}$ Dimensions}

\author{Florian K{\"u}hnel}
	\email{florian.kuhnel@fysik.su.se}
	\affiliation{The Oskar Klein Centre for Cosmoparticle Physics,
			Department of Physics,
			Stockholm University,
			AlbaNova,
			106\.91 Stockholm,
			Sweden}

\author{Bo Sundborg}
	\email{bo.sundborg@fysik.su.se}
	\affiliation{The Oskar Klein Centre for Cosmoparticle Physics,
			Department of Physics,
			Stockholm University,
			AlbaNova,
			106\.91 Stockholm,
			Sweden}

\date{\formatdate{\day}{\month}{\year}, \currenttime}

\begin{abstract}
\noindent [Remark: Much more natural and important models are studied in arXiv:1405.2083. Furthermore, the physical picture given there is much more complete.] The quantum N-portrait -- black holes pictured as Bose-Einstein condensates -- is studied for a special class of models in arbitrary dimensions. In the presence of extra dimensions, the depletion rate is shown to be significantly enhanced in these models as compared to standard semi-classical results. Consequences for large as well as micro condensates are discussed. Speculations are made on a possible connection of higher-dimensional graviton condensates, strings and branes.
\end{abstract}


\maketitle

\section{Introduction}

Recently, Dvali and Gomez proposed a microscopic picture of geometry, and in particular of black holes, in which those can be described as Bose-Einstein condensates (BECs) of soft gravitons \cite{Dvali:2012wq, Dvali:2, Dvali:2013eja, Dvali:3}. Such condensates are self-sustained leaky bound states, and Hawking radiation can be simply understood as quantum depletion of gravitons, which also elegantly explains the negative black holes heat capacity. Moreover, black holes are maximally packed, which means that their occupation number, $N$, is their sole characteristic. Furthermore, it has been conjectured in \cite{Dvali:2} that they are at a critical point of quantum phase transition, which in particular means that quantum effects can never be ignored. These and other findings are promising new insights into the quantum nature of black holes, and eventually of geometry. Along those lines, in Ref.~\cite{Dvali:2013eja}, this approach has been generalized to anti de Sitter and de Sitter space. As far as depletion properties of black holes are concerned, the consequences of this so-called $N$-portrait have essentially only been studied in the $(3 + 1)$-dimensional case. The investigation of precisely those, for a toy model with a dimensionless coupling constant that scales in any dimensions as the General Relativity one in $( 3 + 1 )$ dimensions, for general dimensionality and asymptotically flat space, shall be the focus of this letter.

\section{$\boldsymbol{N}$-portrait in $\boldsymbol{d}$ dimensions}

We will ignore all irrelevant prefactors and set $\hslash = c = 1$. Assuming that the dimensionless self-coupling $\alpha$ is momentum/wavelength ($\lambda$) dependent and is given by
\begin{align}
	\alpha
		&=
								\frac{ L_{*}^{2} }{ \lambda^{2} }								
								\; .
								\label{eq:alpha-d-dim}
\end{align}
The constant $L_{*}$ has dimensionality of length. In order to focus on the behavior of the number $N \gg 1$ of constituents under consideration, we will use Planck units and set $L_{*} = 1$. The Newton potential in $d$ spatial dimensions reads
\begin{align}
	V( r )
		&=
								-
								\frac{ \alpha }{ r^{d - 2} }
								\; .
								\label{eq:V-Newton-d-dim}
\end{align}

Now, following Ref.~\cite{Dvali:2}, we ask for the $N$-dependence of the wavelength $\lambda( N )$ for which the constituents form a self-sustained bound state of size $L \sim \lambda \gg 1$. The corresponding equilibrium condition can be obtained from equating the kinetic energy that each single quanta has, $E_{L} = 1 / L$, with the energy of the collective binding potential \eqref{eq:V-Newton-d-dim},
\begin{align}
	\frac{ 1 }{ L }
		=
	E_{L}
		&\overset{!}{=}
								V
		\simeq
								N\.\frac{ \alpha }{ L^{d - 2} }
		\simeq
								N\.\frac{ 1 }{ L^{d} }
								\; .
								\label{eq:}
\end{align}
Thus, we find
\begin{subequations}
\begin{align}
	L
		&\simeq
								N^{\frac{ 1 }{ d - 1 }}
								\; .
								\label{eq:L(N)}
\end{align}
Hence, the number of quanta grows a generalized area law, $N \sim L^{d - 1}$. Furthermore, we have
\vs{-1mm}
\begin{align}
	\alpha
		&\simeq
								N^{- \frac{ 2 }{ d - 1 }}
								\; .
								\label{eq:alpha(N)}
\intertext{The mass $M$ of the condensate, which is approximately given by the sum of the constituent's energies, reads\vs{-4mm}}
	M
		&\simeq
								N^{\frac{ d - 2 }{ d - 1 }}
								\; .
								\label{eq:M-d-dim}
\end{align}
\end{subequations}

We estimate the depletion rate, $\Gamma$, as follows. The idea is that any interactions among constituents will scatter at least one quantum out of the bound condensate, causing depletion. Focusing on $2 \ra 2$ processes, this rate essentially consist of three factors: one that counts all possible pairs, \ie~$N( N - 1 ) \simeq N^{2}$, then $\alpha^{2} \equiv \alpha( N )^{2}${\,---\,}since it is a two-vertex scattering{\,---}, and furthermore the characteristic energy, $E( N ) \sim L( N )^{-1} \sim N^{- 1 / ( d - 1 )}$, of the depletion process. Hence,
\vs{-2mm}
\begin{align}
	\Gamma
		&\sim
								N^{2}\.\alpha( N )^{2}\.E( N )
		\sim
								N^{\frac{ 2 d - 7 }{ d - 1 }}
								\; .
								\label{eq:Gamma-d-dim}
\end{align}
Let $\Delta M = 1 / L \sim N^{- 1 / ( d - 1 )}$ be the energy of one emitted quantum. Then the above rate translates to the emission rate
\begin{align}
	\frac{ \d M }{ \d t }
		&=
								-\,
								\Delta M\;\Gamma
		\sim
								N^{\frac{ 2 d - 8 }{ d - 1 }}
		\sim
								M^{\frac{ 2 d - 8 }{ d - 2 }}
								\; .
								\label{eq:dM/dt-in-terms-of-N-d-dim}
\end{align}
As expected, for three spatial dimensions, this agrees with the simple standard semi-classical/thermo-dynamic result (\cf~\eg~\cite{Argyres:1998qn})\footnote{This result applies in the case in which the Schwarzschild radius is much smaller than the extent of the\\[-1mm] extra-dimension, which will always be assumed in the reminder of the work.\\[-4.8mm]}
\begin{align}
	\frac{ \d M }{ \d t }
		&\sim
								\big[
									\text{Area}( M )
								\big]\;
								T_{\text{Hawking}}^{d + 1}( M )
		\sim
								M^{- \frac{ 2 }{ d - 2 }}
								\; .
								\label{eq:dM/dt-Hawking}
\end{align}
However, for all $d \ne 3$ there are substantial differences.

In \cite{Cardoso:2005vb} the energy loss of gravitons into extra-dimensions has been estimated numerically in a more sophisticated way, by precisely investigating all the greybody factors. The set-up therein is a brane-world scenario where gravity is allowed to propagate into the non-compact bulk, while all the (standard model) particles are confined into a $(3 + 1)$-dimensional brane. Concretely\footnote{Numbers are taken from Table III of Ref.~\cite{Cardoso:2005vb}. Each rate 
has a numerical accuracy of about $5\%$.\\[-4.8mm]}
\vs{-2mm}
\begin{align}
	\frac{ \dot{M}\big|_{\text{graviton}} }{ \dot{M}\big|_{\text{scalar}} }\,( d = 3,\,4,\,5,\,6,\,7,\,8 )
		&\approx
							 	0.02,\,0.2,\,0.6,\,0.91,\,1.9,\,2.5	
								\; ,
								\label{eq:Mdot-graviton/Mdot-scalar-numerical-vaules}
\end{align}
where these numbers are per degree of freedom, and a dot stands for a time derivative, and{\,---\,}as indicated{\,---\,}these number are compared to the scalar (\eg~Higgs) emission into the respective dimension.

To compare the above numerical results to our analytical estimate \eqref{eq:dM/dt-in-terms-of-N-d-dim}, we define a function which isolates the exponent of $\dot{M}$. Such a function is
\begin{align}
	\Delta( d )
		&\defas
								\frac{ \log\!
								\Big[
									\frac{ \dot{M}( d ) }{ \dot{M}( d_{1} ) }
								\Big] }
								{ \log\!
								\Big[
									\frac{ \dot{M}( d_{2} ) }{ \dot{M}( d_{1} ) }
								\Big] }
								\; ,
								\label{eq:}
\end{align}
which is a measure of the growth and independent of $M$, or $N$, respectively. Above, $d_{1} \ne d_{2}$ are two reference dimensions; We will set $d_{1} = 3$ and $d_{1} = 4$ in the remainder of this work. However, the precise values are not important.

\begin{figure}
	\centering
	\includegraphics[scale=0.52]{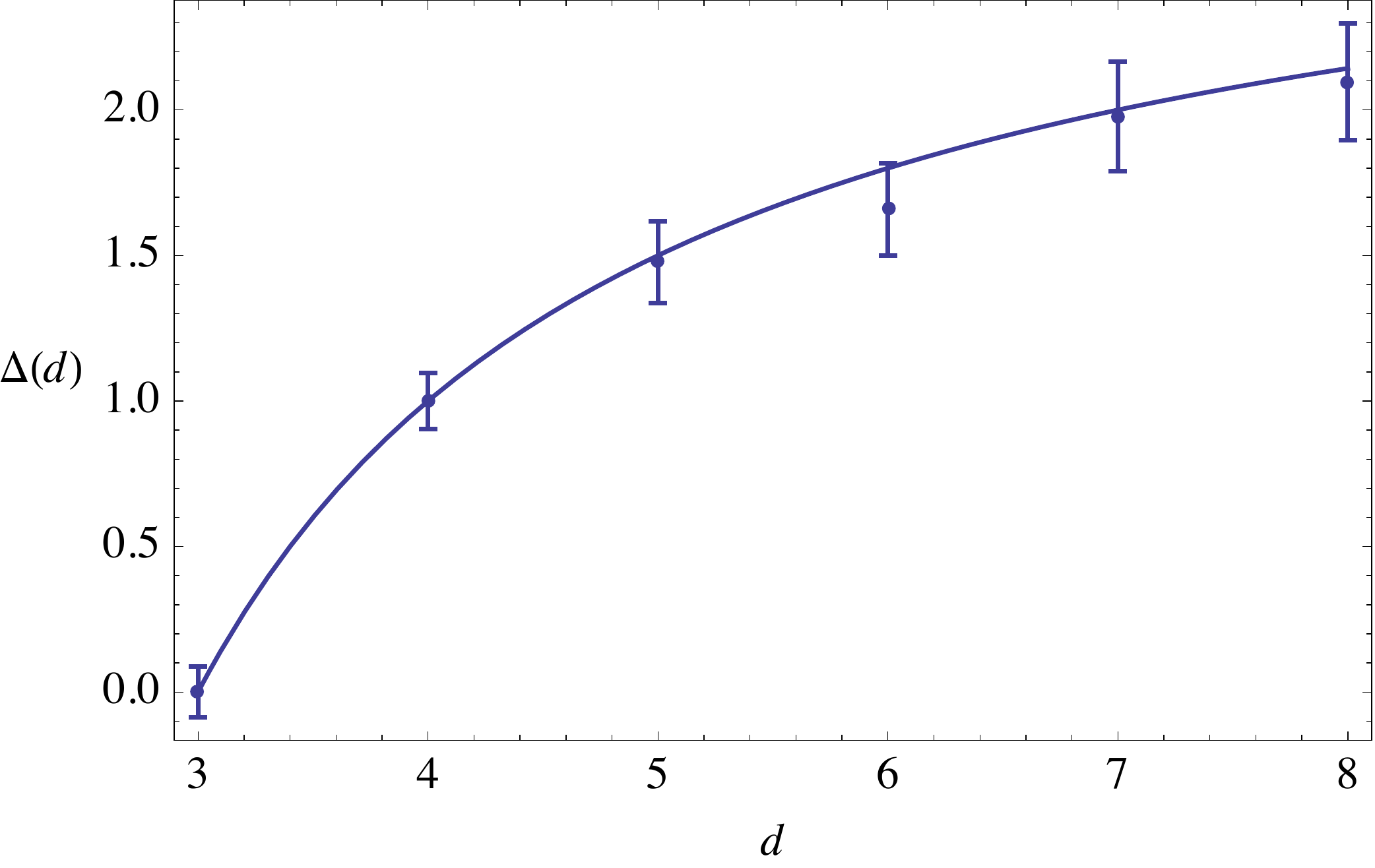}
	\caption{$\Delta$ as a function of spatial dimension $d$ (solid line);
			Data points are taken from Ref.~\cite{Cardoso:2005vb}.}
	\label{fig:Graviton-Emission}
\end{figure}

Fig.~\ref{fig:Graviton-Emission} depicts the function $\Delta( d )$ for spatial dimensions $d$ in the interval $[ 3, 8 ]$. We clearly see a remarkable agreement of the numerical results of \cite{Cardoso:2005vb} with our analytical estimate Eq.~\eqref{eq:dM/dt-in-terms-of-N-d-dim}. Now, a power-law $\dot{M} \sim N^{\theta( d )}$ yields
\begin{align}
	\Delta( d )
		&=
								\frac{ \theta( d ) - \theta( d_{1} ) }
								{ \theta( d_{2} ) - \theta( d_{1} ) }
								\; ,
								\label{eq:}
\end{align}
which has special values $\Delta( d_{1} ) = 0$ and $\Delta( d_{2} ) = 1$.\footnote{$\Delta$ actually is a M{\"o}bius transformations if $\theta$ is one.} We find that both rates Eqs.~(\ref{eq:dM/dt-in-terms-of-N-d-dim}, \ref{eq:dM/dt-Hawking}) give rise to the same $\Delta( d )$! Although their specific power-law dependence is different, they fall{\,---\,}in some sense{\,---\,}into the same universality class. It will be interesting to explore this relation further \cite{KS}.

The rate \eqref{eq:Gamma-d-dim} can easily be integrated to give the half-life time, $\tau$, of the $d$-dimensional black-hole-like object. Here we define this time as the time at which half of the constituents is depleted away, which yields\footnote{Note that for $d = 6$ the $N$-dependence drops out.}
\vs{-3mm}
\begin{align}
	\tau
		&\sim
								N^{\frac{ 5 }{ d - 1 } - 1 }
								\; .
								\label{eq:}
\end{align}
This result is in units of Planck time, $\tau_{\text{Planck}} \approx 10^{-43}\.{\rm s}$, and has to be compared to the age of the Universe, $\tau_{\text{H}} \approx 10^{17}\.{\rm s} \approx 10^{60}\.\tau_{\text{Planck}}$. Setting $N \equiv 10^{\theta}$, we find
\begin{align}
	\frac{ \tau }{ \tau_{\text{H}} }
		&\sim
								\Ocal( 1 ) \cdot 10^{ \frac{ 5 \theta }{ d - 1 } - \theta - 60 }
								\; .
								\label{eq:}
\end{align}

Table \ref{tab:Lifetimes} shows hypothetical half-life time estimates for various astrophysical objects in units of the age of the Universe. We see that even super-heavy objects do not have a half-life time close $\tau_{\text{H}}$. Hence, for large extra dimensions\footnote{Here, an extra-dimension is referred to as large if its extend is much larger than $L$.}, essentially all astro-physical graviton-like BECs are relatively short-lived.

\begin{table}
\begin{center}
\vs{2mm}
\begin{tabular}{|l||l|l|l|l||l|}
\hline
\backslashbox{\;$d$}{}	&\;$\overset{\text{Moon}\vphantom{_{_{_{_{_{}}}}}}}{\approx 10^{23}\.{\rm kg}}
						\vphantom{1^{^{^{^{^{^{^{^{^{1}}}}}}}}}}$\;				
					&\;$\overset{\text{Earth}\vphantom{_{_{_{_{}}}}}}{\approx 10^{25}\.{\rm kg}}$\;
					&\;$\overset{\text{Sun}\vphantom{_{_{_{_{_{}}}}}}}{\approx 10^{30}\.{\rm kg}}$\;
					&\;$\overset{\text{Srg\.A$^{\!*}$}\vphantom{_{_{_{_{}}}}}}{\approx 10^{36}\.{\rm kg}}$\;
					&\;$\overset{\text{Hubble patch}\vphantom{_{_{_{_{_{}}}}}}}{\approx 10^{53}\.{\rm kg}}$\;\\
\hline\hline
	\;3				&\;$10^{+33}\.\tau_{\text{H}}\vphantom{_{_{_{_{1}}}}}\vphantom{^{^{^{^{.}}}}}$
					&\;$10^{+39}\.\tau_{\text{H}}\vphantom{_{_{_{_{1}}}}}\vphantom{^{^{^{^{.}}}}}$
					&\;$10^{+54}\.\tau_{\text{H}}\vphantom{_{_{_{_{1}}}}}\vphantom{^{^{^{^{.}}}}}$
					&\;$10^{+72}\.\tau_{\text{H}}\vphantom{_{_{_{_{1}}}}}\vphantom{^{^{^{^{.}}}}}$
					&\;$10^{+123}\.\tau_{\text{H}}\vphantom{_{_{_{_{1}}}}}\vphantom{^{^{^{^{.}}}}}$\\
\hline
	\;4				&\;$10^{-29}\.\tau_{\text{H}}\vphantom{_{_{_{_{1}}}}}\vphantom{^{^{^{1}}}}$
					&\;$10^{-27}\.\tau_{\text{H}}\vphantom{_{_{_{_{1}}}}}\vphantom{^{^{^{1}}}}$
					&\;$10^{-22}\.\tau_{\text{H}}\vphantom{_{_{_{_{1}}}}}\vphantom{^{^{^{1}}}}$
					&\;$10^{-16}\.\tau_{\text{H}}\vphantom{_{_{_{_{1}}}}}\vphantom{^{^{^{1}}}}$
					&\;$10^{+1{\color{white}00}}\.\tau_{\text{H}}\vphantom{_{_{_{_{1}}}}}\vphantom{^{^{^{1}}}}$\\
\hline
	\;5				&\;$10^{-50}\.\tau_{\text{H}}\vphantom{_{_{_{_{1}}}}}\vphantom{^{^{^{1}}}}$
					&\;$10^{-49}\.\tau_{\text{H}}\vphantom{_{_{_{_{1}}}}}\vphantom{^{^{^{1}}}}$
					&\;$10^{-47}\.\tau_{\text{H}}\vphantom{_{_{_{_{1}}}}}\vphantom{^{^{^{1}}}}$
					&\;$10^{-45}\.\tau_{\text{H}}\vphantom{_{_{_{_{1}}}}}\vphantom{^{^{^{1}}}}$
					&\;$10^{-40{\color{white}0}}\.\tau_{\text{H}}\vphantom{_{_{_{_{1}}}}}\vphantom{^{^{^{1}}}}$\\
\hline
	\;6				&\;$10^{-60}\.\tau_{\text{H}}\vphantom{_{_{_{_{1}}}}}\vphantom{^{^{^{1}}}}$
					&\;$10^{-60}\.\tau_{\text{H}}\vphantom{_{_{_{_{1}}}}}\vphantom{^{^{^{1}}}}$
					&\;$10^{-60}\.\tau_{\text{H}}\vphantom{_{_{_{_{1}}}}}\vphantom{^{^{^{1}}}}$
					&\;$10^{-60}\.\tau_{\text{H}}\vphantom{_{_{_{_{1}}}}}\vphantom{^{^{^{1}}}}$
					&\;$10^{-60{\color{white}0}}\.\tau_{\text{H}}\vphantom{_{_{_{_{1}}}}}\vphantom{^{^{^{1}}}}$\\
\hline
\end{tabular}
\end{center}
\caption{Approximate half-life times of certain astrophysical objects in units of the age of the Universe for various spatial dimensions. Here the $( 3 + 1 )$-dimensional Planck mass has been used.}
\label{tab:Lifetimes}
\end{table}

A rather different situation to the one studied above is constituted by micro black holes. Here we take the extra dimensions to be compactified, which, for illustrative purpose, will be such that the higher-dimensional Planck mass is equal to $1\.{\rm MeV}$. Then, a gravitational bound state of $10\.{\rm TeV}$ still contains $N \sim 10^{7\.\frac{ d - 1 }{ d - 2 }} \gg 1$ quanta [\cf~Eq.~\eqref{eq:M-d-dim}], and one finds
\begin{align}
	\tau_{10\.{\rm TeV}}( d = 4,\,5\,,6 )
		&\sim
								10^{-13},\,
								10^{-18},\,
								10^{-20}
								\;{\rm s}
								\; ,
								\label{eq:}
\end{align}
which means that these objects are extremely short-lived.\\

\section{Summary \& Speculation}
\label{sec:Summary-and-Speculation}

In the work we have investigated $d$-dimensional condensates of soft hypothetical quanta, which have a dimensionless coupling constant that scales in any dimensions as the General Relativity one in $( 3 + 1 )$ dimensions. 
We estimated the depletion rate and found that for $d > 3$ it is significantly enhanced as compared to the standard result, based on semi-classical assumptions. We calculated the half-life time of astrophysical, as well as micro graviton-like condensates. It turns out that in the presence of additional dimensions, those objects deplete much more rapid.

If the higher-dimensional BECs are less long-lived than the higher-dimensional Schwarz-schild solutions, they should be identified with other objects. Ultraviolet problems in gravity are potentially worse in higher dimensions and it might be instructive to glance at theories that are well-defined in the UV. In string theory, which is UV regular, we typically expect large strings to stretch outside the Schwarzschild radius in higher dimensions. This is because the growth of the typical string size with energy is independent of dimension, while the Schwarzschild radius grows slower and slower with energy in higher dimensions. Such a fundamental string can be regarded as a solitonic solution \cite{solitonic-solution} to an ultraviolet completion of a higher-dimensional gravity theory, and it could be represented as a Bose-Einstein condensate. We do not view the dictionary from the BEC picture to spacetime as an obvious one, and other candidates than black holes might be relevant. The ultimate test comes from matching of calculable physical properties.



One test would be to compare lifetimes. Although decays of massive strings are much more complicated than black holes, because the spectrum of string solutions is exponentially rich at high mass, some results exist. In particular, average decay rates have been studied, \cf~\cite{average-decay-rates} and references therein. These results also do not match the standard black hole result, and indeed show a higher decay rate, albeit not the same as the Bose-Einstein condensate picture yields. However, the mentioned results are only perturbative and to first order. The higher-order corrections could shed light on this question.


However, we have no particular reason to match the condensates uniquely to average string solutions, as there typically are many other classes of gravitational solitons, like D-branes, in string theories. Furthermore, each of these solutions comes with many possible oscillatory excitations. The point here is that there are other solutions than black holes that may correspond to the higher-dimensional Bose-Einstein condensates.

Although the classical interpretation of higher-dimensional BECs is not yet fully clear, we believe that they will have important physical consequences.


\acknowledgements
It is a pleasure to thank Gia Dvali, Daniel Flassig, Stefan Hofmann, Edvard M{\"o}rtsell, Alexander Pritzel, and Nico Wintergerst for stimulating discussions and important remarks. This work was supported by the Swedish Research Council (VR) through the Oskar Klein Centre.


\end{document}